\documentclass[aps,prb,reprint]{revtex4-2}
\newcommand{\degree}{^\circ}
\usepackage{amsmath}
\usepackage{amssymb}
\usepackage{graphicx}
\usepackage{multirow}

\begin{document}
\title{Phase competition and negative piezoelectricity in interlayer-sliding ferroelectric ZrI$_2$}

\author{Ning Ding}
\author{Jun Chen}
\author{Churen Gui}
\author{Haipeng You}
\author{Xiaoyan Yao}
\email{Email: yaoxiaoyan@seu.edu.cn}
\author{Shuai Dong}
\email{Email: sdong@seu.edu.cn}
\affiliation{School of Physics, Southeast University, Nanjing 21189, China}
\date{\today}

\begin{abstract}
The so-called interlayer-sliding ferroelectricity was recently proposed as an unconventional route to pursuit electric polarity in van der Waals multi-layers, which was already experimentally confirmed in WTe$_2$ bilayer even though it is metallic. Very recently, another van der Waals system, i.e., the ZrI$_2$ bilayer, was predicted to exhibit the interlayer-sliding ferroelectricity with both in-plane and out-of-plane polarizations [Phys. Rev. B \textbf{103}, 165420 (2021)]. Here the ZrI$_2$ bulk is studied, which owns two competitive phases ($\alpha$ \textit{vs} $\beta$), both of which are derived from the common parent $s$-phase. The $\beta$-ZrI$_2$ owns a considerable out-of-plane polarization ($0.39$ $\mu$C/cm$^2$), while its in-plane component is fully compensated. Their proximate energies provide the opportunity to tune the ground state phase by moderate hydrostatic pressure and uniaxial strain. Furthermore, the negative longitudinal piezoelectricity in $\beta$-ZrI$_2$ is dominantly contributed by the enhanced dipole of ZrI$_2$ layers as a unique characteristic of interlayer-sliding ferroelectricity, which is different from many other layered ferroelectrics with negative longitudinal piezoelectricity like CuInP$_2$S$_6$.
\end{abstract}
\maketitle

\section{Introduction}
Ferroelectrics with switchable electric polarizations have attracted great attentions for their unique physical properties and wide applications, such as micro-electro-mechanical systems and nonvolatile memories \cite{Scott-science-2007,Scott-science-1989,Dawber-Rev-Mod-Phys-2005,uchino1996piezoelectric}. New applications such as ferroelectric tunnel junctions and ferroelectric photovoltaics have also been proposed in recent years \cite{Zhuravlev-PRL-2005, Yuan-Nat-Mater-2011}. In addition to conventional proper ferroelectricity originated from $d^0$ ions, $6s^2$ lone pairs, or polar groups, improper ferroelectricity has been intensively studied in the past two decades, whose polarizations can be originated from charge ordering, magnetism, and even non-polar phonons \cite{Cheong-Nat-Mater-2007}. These new branches not only extend the scope of ferroelectrics but also provide new functions for applications \cite{Dong:Ap,Dong:Nsr}.

As an emergent branch of polar materials, two-dimensional (2D) ferroelectrics were highly concerned in recent years \cite{Wumenghao-wires-2018,guan-AEM-2020,An-APL-Mater-2020}, which are mostly derived from van der Waals (vdW) layered materials, such as In$_2$Se$_3$ \cite{Ding-Nat-Com-2017}, and CuInP$_2$S$_6$ \cite{Liu-Nat-Com-2016}. These 2D ferroelectrics not only own promising future for integratability of nanoscale devices \cite{Yuan-Nat-Com-2019}, but also provide a unique platform to explore unconventional physical mechanisms of polarity, such as noncollinear ferrielectricity \cite{Lin-PRL-2019}, as well as various magnetic ferroelectrics \cite{Ding-PRB-2020,J.J.Zhang-JACS-2018}.

In 2017, Wu \textit{et al.} proposed the concept of interlayer sliding ferroelectricity, namely that out-of-plane ferroelectric polarizations could be induced by interlayer sliding in vdW bilayers such as $h$-BN, AlN, etc. \cite{li-ACS-Nano-2017}. Soon, this mechanism was adopted to explain the ferroelectricity observed in WTe$_2$ bilayer \cite{Fei-Nature-2018,Yang-JPCL-2018,Liu-Nanoscale-2019}, although it is a topological semimetal. As a newly developed sub-branch of unconventional polarity, more 2D materials in this category are expected to be explored, as well as their unique physical properties.

Very recently, the interlayer sliding ferroelectricity in ZrI$_2$ bilayer was theoretically studied \cite{Zhang-PRB-bilayer-ZrI2-2021}, which predicted both the in-plane and out-of-plane polarizations. However, the experimental verification of bilayer system needs very precise experimental fabrication and measurements, which are technically challenging.

In this work, the ZrI$_2$ bulk will be theoretically studied, whose interlayer sliding ferroelectricity is similar but not identical to its bilayer counterpart. In experiment, two polymorph forms of ZrI$_2$ bulk were synthesized, which even coexisted \cite{corbett1982second}. The monolinic $\alpha$-ZrI$_2$ (space group $P2_1/m$) is isostructural with $1T'$-MoTe$_2$/WTe$_2$ \cite{Brown-Crystallographica-1966,Singh-PRL-2020,Huang-NC-2019}, while the orthorhombic $\beta$-ZrI$_2$ (space group $Pmn2_1$) is isostructural with $T_d$-MoTe$_2$/WTe$_2$ \cite{Brown-Crystallographica-1966,Singh-PRL-2020,Huang-NC-2019}. Our work will clarify the relationship of these phases: the parent phase $s$-ZrI$_2$ (equivalent to $T_0$-MoTe$_2$/WTe$_2$ phase \cite{Singh-PRL-2020,Huang-NC-2019}), and the derived phases: non-polar $\alpha$-ZrI$_2$ and polar $\beta$-ZrI$_2$. Furthermore, an intriguing negative longitudinal piezoelectric effect is revealed in this vdW material, which not only arises from the volume reduction of vdW layer upon pressure or compressive strain, but also mostly origins from the strengthened electric dipole in ZrI$_2$ layer, as a unique characteristic of its ferroelectric origin from interlayer sliding.

\begin{figure}
\includegraphics[width=0.48\textwidth]{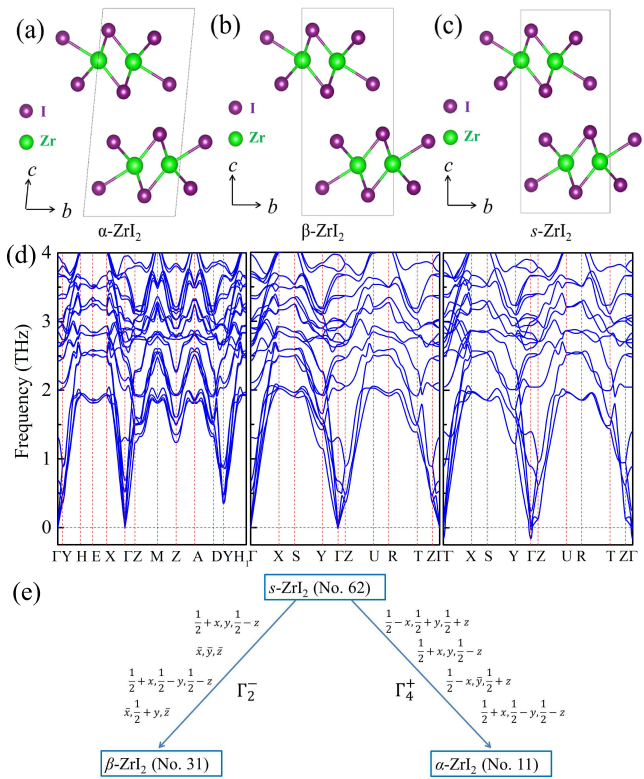}
\caption{(a-c) Schematic structures (side view) of various phases of ZrI$_2$ bulk. Each unit cell (u.c.) contains two vdW layers and four Zr's. (d) The corresponding phonon spectra. No imaginary frequency exists for the $\alpha$- and $\beta$-phases, while weak imaginary frequencies exist from $\Gamma$ to $Z$ for the $s$-ZrI$_2$, implying its dynamic instability. The corresponding Brillouin zones are shown in Fig.~S1 of SM \cite{sm}. (e) The evolution tree of these three phases. Starting from the high symmetric space group of parent phase, the space groups of $\alpha$- and $\beta$-phases can be obtained by reducing four symmetric operations, respectively.}
\label{fig1}
\end{figure}

\section{Methods}
The first-principles calculations based on density functional theory (DFT) were performed with the projector augmented wave (PAW) pseudopotentials as implemented in Vienna {\it ab initio} Simulation Package (VASP) \cite{PAW-PRB-1996}. The Perdew-Burke-Ernzerhof (PBE) parameterization of generalized gradient approximation (GGA) was used for the exchange-correlation functional \cite{PBE-PRL-1996}. The plane-wave cutoff energy was $500$ eV. The \emph{k}-point grids of $11\times6\times3$ is adopted for both structural relaxation and static computation. To describe the interlayer interaction, the vdW correction for potential energy was implemented using the DFT-D2 method \cite{D2-JCC-2006}. The convergent criterion for the energy was set to $10^{-6}$ eV, and that of the Hellman-Feynman forces during the structural relaxation was $0.001$ eV/\AA. In addition, the Heyd-Scuseria-Ernzerhof (HSE06) functional is also adopted for comparison of band gaps \cite{HSE06}.

The phonon spectra were calculated using the finite differences method to verify the dynamic stability of structures \cite{Finite-differences-PRB-2002,Finite-difference-PRB-2005}. The polarization was calculated using the standard Berry phase method \cite{Polarization-PRB-1993,Resta-RevModPhys-1994}, and the possible ferroelectric switching paths were obtained by linear interpolation of two end states. In addition, the density functional perturbation theory (DFPT) was employed to calculate the piezoelectric coefficients \cite{DFPT-PRB-1997}.

\section{Results \& Discussion}
\subsection{Crystalline \& electronic structures}
Figures~\ref{fig1}(a-b) show the structures of $\alpha$-ZrI$_2$ (monoclinic, space group No. 11 $P2_1/m$) and $\beta$-ZrI$_2$ (orthorhombic, space group No. 31 $Pmn2_1$), which look quite similar but with different monoclinic angle and interlayer sliding. Their DFT optimized lattice constants agree well with the experimental values, as compared in Table~\ref{table1}, which ensure the accuracy of following calculations. The DFT atomic positions for different phases are summarized in Table S1 of Supplemental Materials (SM) \cite{sm}.

\begin{table}
\caption{Basic physical properties of three phases of ZrI$_2$ bulk from DFT calculations and experiments (Exp). The lattice constants are in unit of \AA, the energies ($E$) relative to the $\beta$-phase are in unit of meV/u.c., and the band gaps are in unit of eV. $\angle$ denotes the monoclinic angle, in unit of $\degree$. The in-plane lattice constants of $\alpha$/$\beta$/$s$-phases obtained in our DFT relaxation are almost identical.}
\centering
\begin{tabular*}{0.48\textwidth}{@{\extracolsep{\fill}}lcccccccc}
\hline \hline
	&  S.G. & $a$ & $b$ & $c$ & $\angle$ & $E$ & gap \\
\hline
$\alpha$-ZrI$_2$ (DFT) & \multirow{2}{*}{$P2_1/m$} & $3.749$ & $6.865$ & $14.881$ & $84.05$ & $1.1$ & $0.11$ \\
$\alpha$-ZrI$_2$ (Exp \cite{guthrie1981synthesis}) &  & $3.741$ & $6.821$ & $14.937$ & $84.34$ & - & $\sim0.1$\\
$\beta$-ZrI$_2$ (DFT) & \multirow{2}{*}{$Pmn2_1$} & $3.749$ & $6.865$ & $14.801$ & $90$ & $0$ & $0.15$ \\
$\beta$-ZrI$_2$ (Exp \cite{corbett1982second}) &  & $3.744$ & $6.831$ & $14.886$ & $90$ & - & - \\
$s$-ZrI$_2$ (DFT) & $Pnma$ & $3.749$ & $6.865$ & $14.808$ & $90$ & $28$ & $0.20$ \\
\hline \hline
\end{tabular*}
\label{table1}
\end{table}

As expected, the DFT energies of $\alpha$-ZrI$_2$ and $\beta$-ZrI$_2$ are very close: only $\sim1.1$ meV/u.c. higher for $\alpha$-ZrI$_2$. Such a small energy difference can well explain the experimental observation that the monoclinic $\alpha$-phase and orthorhombic $\beta$-phase coexisted in the samples \cite{corbett1982second}.

The phonon spectra have also been calculated for these two phases to check their dynamic stability, as displayed in Fig.~\ref{fig1}(d). No imaginary frequency exists for both the $\alpha$- and $\beta$-phases, implying that both of them are dynamically stable.

However, the space group $P2_1/m$ is not the parent group of $Pmn2_1$. Thus, the $\alpha$-phase can not be recognized as the high-symmetric paraelectric phase above the ferroelectric transition temperature. Instead, there are must be another higher symmetric one as the high-temperature paraelectric phase of ZrI$_2$, which was proposed as the $s$-ZrI$_2$ (space group No. 62 $Pnma$, which is the common parent group of both $Pmn2_1$ and $P2_1/m$), as shown in Fig.~\ref{fig1}(c). Its dynamic stability is also checked. As shown in the right panel of Fig.~\ref{fig1}(d), this $s$-phase is dynamically unstable as expected for paraelectric parent phase. There are two imaginary frequencies: one is an unstable optical phonon mode $\Gamma^-_2$ at zone center, corresponding to the interlayer sliding of neighboring layers; and another one is an elastic instability mode $\Gamma^+_4$, causing the shearing distortion of the orthorhombic unit cell \cite{BCS-2011}.
Indeed, the two irreducible representations $\Gamma_2^-$ and $\Gamma_4^+$ transform the $Pnma$ phase to the $Pmn2_1$ and $P2_1/m$, respectively. The energy profiles of these two distortions can be found in Fig.~S2 of SM \cite{sm}, both of which are in the shape of double well. Thus, the evolution tree of these three phases is summarized in Fig.~\ref{fig1}(e): the $s$-ZrI$_2$ is the real high-symmetric paraelectric phase, while the $\alpha$- and $\beta$-phases are two ``sister" phases derived from the $s$-phase.

\begin{figure}
\includegraphics[width=0.48\textwidth]{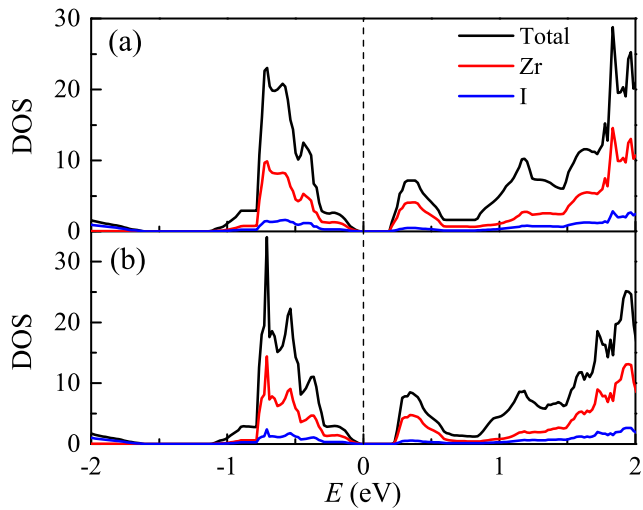}
\caption{Density of states and their atomic projects of (a) $\alpha$-ZrI$_2$ and (b) $\beta$-ZrI$_2$, calculated using pure GGA method.}
\label{fig2}
\end{figure}

The calculated electronic properties indicate that both $\alpha$-ZrI$_2$ and $\beta$-ZrI$_2$ are semiconductors with band gaps $\sim110$ meV and $150$ meV respectively, as shown in Fig.~\ref{fig2}. Except the tiny difference of band gaps, their electronic structures are quite similar, and the band gaps agree with the experimental one ($\sim0.1$ eV \cite{guthrie1981synthesis}).

The HSE calculations can lead to larger band gaps $\sim520$ meV for $\alpha$-ZrI$_2$ and $\sim560$ meV for $\beta$-ZrI$_2$, respectively, much higher than the experimental one \cite{guthrie1981synthesis}. Noting that the experimental band gap might also be underestimated due to defect levels in forbidden band of semiconductors \cite{dong:prl14,NPG-ASIA-MAT}. Therefore, further experiments are encouraged to verify their intrinsic band gaps.

The spin-orbit coupling (SOC) effects are also taken into consideration (Fig.~S3 in SM \cite{sm}), considering the existence of heavy element iodine. Even though, their electronic structures are insensitive to the SOC effect, because the electronic states around the Fermi level are mainly from Zr$^{2+}$ instead of I$^-$.

\subsection{Ferroelectricity of $\beta$-ZrI$_2$}
Using the standard Berry phase calculation, the electric polarization of $\beta$-ZrI$_2$ is estimated as $0.39$ $\mu$C/cm$^2$, pointing along the $c$-axis, which is much larger than many others in the same category, e.g. $0.03$ $\mu$C/cm$^2$ in WTe$_2$ bilayer \cite{Liu-Nanoscale-2019,Yang-JPCL-2018}, but smaller than $h$-BN bilayer ($0.68$ $\mu$C/cm$^2$) \cite{li-ACS-Nano-2017}.

\begin{figure}
\includegraphics[width=0.48\textwidth]{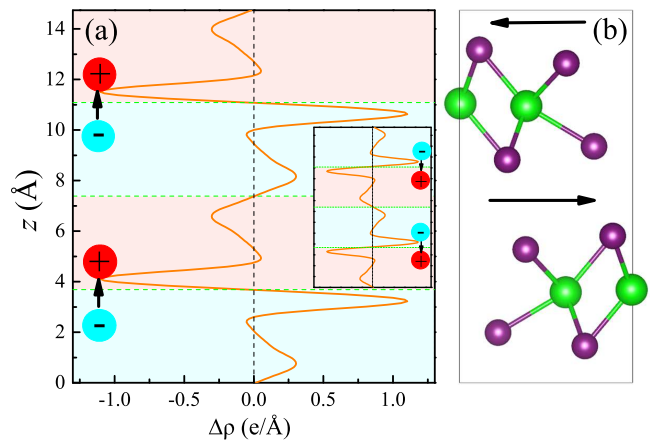}
\caption{Origin of interlayer-sliding ferroelectricity in $\beta$-ZrI$_2$. (a) The planar-averaged differential charge density ($\Delta\rho$) along the $c$ axis (characterized by the height $z$), with $s$-ZrI$_2$ as the reference. Insert: the opposite polarization. Here the positive value of $\Delta\rho$ represents more electrons while its negative value means more holes. The red and cyan spheres represent the charge centers of positive and negative regions (distinguished by colors), respectively. The arrows denote the dipoles. (b) The corresponding structure of (a) for comparison. The in-plane component of polarization (indicated by arrows) can exist in bilayer but is cancelled in bulk for its antiferroelectric ordering between neighboring bilayers.}
\label{fig3}
\end{figure}

To clarify the physical origin of its polarization, the planar-averaged differential charge density is calculated as a function of height along the $c$-axis \cite{C5CP00011D}, with the paraelectric $s$-ZrI$_2$ as the reference base. For the $\beta$-ZrI$_2$, a charge dipole is created in each ZrI$_2$ layer, as shown in Fig.~\ref{fig3}(a). In other words, the interlayer sliding breaks the inversion symmetry between upper and lower Zr-I bonds (i.e., their chemical environments), making one side more charge positive while the opposite side more negative. Such a polarization can be reversed by further interlayer sliding, as shown in the insert of Fig.~\ref{fig3}(a).

It should be noted that both the out-of-plane and in-plane polarizations were predicted in ZrI$_2$ bilayer \cite{Zhang-PRB-bilayer-ZrI2-2021}, different from the $\beta$-ZrI$_2$ bulk which only owns the out-of-plane polarization. First, our DFT calculation indeed confirms the existence of in-plane polarization in bilayer, the details can be shown in Fig. S4 of SM \cite{sm}. Second, such an in-plane polarization is canceled between nearest-neighbor bilayers in bulk, because the A-B stacking and B-A stacking lead to an antiferroelectric ordering of in-plane polarization, as illustrated in Fig.~\ref{fig3}(b).

\begin{figure}
\includegraphics[width=0.48\textwidth]{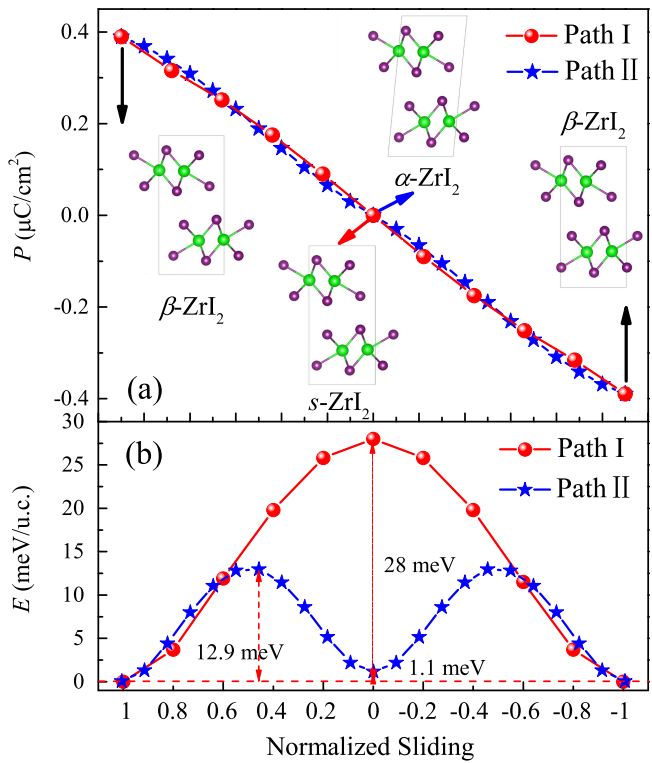}
\caption{(a) Two possible switching paths of polarization ($P$) in $\beta$-ZrI$_2$. The $s$-ZrI$_2$ is the intermediate state of path I, while the $\alpha$-ZrI$_2$ is the intermediate state of path II. Both of these paths are achieved via the interlayer sliding of vdW layers, while the path II also includes the monoclinic angle distortion of cell. The interlayer sliding is normalized to its optimized value. (b) The corresponding energy barriers of these two switching paths for $\beta$-ZrI$_2$.}
\label{fig4}
\end{figure}

The ferroelectric switching of $\beta$-ZrI$_2$ is also nontrivial, as illustrated in Fig.~\ref{fig4}. Both the parent phase $s$-ZrI$_2$ and sister phase $\alpha$-ZrI$_2$ can play as the intermediate states, leading to two different switching paths (I \textit{vs} II). On one hand, these two paths show almost identical behavior of polarization changing as a function of normalized interlayer sliding along the $b$-axis, as shown in Fig.~\ref{fig4}(a). On the other hand, these paths own totally different switching barriers, as shown in Fig.~\ref{fig4}(b). The more conventional path I, which takes $s$-ZrI$_2$ as the intermediate phase, has a higher energy barrier $\sim28$ meV/u.c. (i.e., $7$ meV/f.u.), which is unimodal since the $s$-ZrI$_2$ is dynamic unstable. This barrier is higher than that of ZrI$_2$ bilayer \cite{Zhang-PRB-bilayer-ZrI2-2021}, mainly because that the vdW coupling is double side in bulk but unilateral in bilayer. In contrast, the dynamic stable $\alpha$-ZrI$_2$ as the intermediate state leads to double peaks barrier $\sim12.9$ meV/u.c. (i.e., $3.2$ meV/f.u.) for the path II. Therefore, the path II may be the more preferred one in real switching process, with synchronous monoclinic distortion and interlayer sliding, which is beyond previous studies of interlayer sliding ferroelectrics \cite{Liu-Nanoscale-2019,Yang-JPCL-2018,li-ACS-Nano-2017}

\subsection{Mechanical tuning}
The energy proximity between $\alpha$-ZrI$_2$ and $\beta$-ZrI$_2$ provides an opportunity to tune their competition via external stimulations. Besides the electrical switching as discussed in above subsection, the mechanical approaches can also be available.

First, the aforementioned path II has already demonstrated that the monoclinic distortion during the ferroelectric switching. As an inverse effect, a shear strain may drive the $\beta$-ZrI$_2$ to $\alpha$-ZrI$_2$.

\begin{figure}
\includegraphics[width=0.48\textwidth]{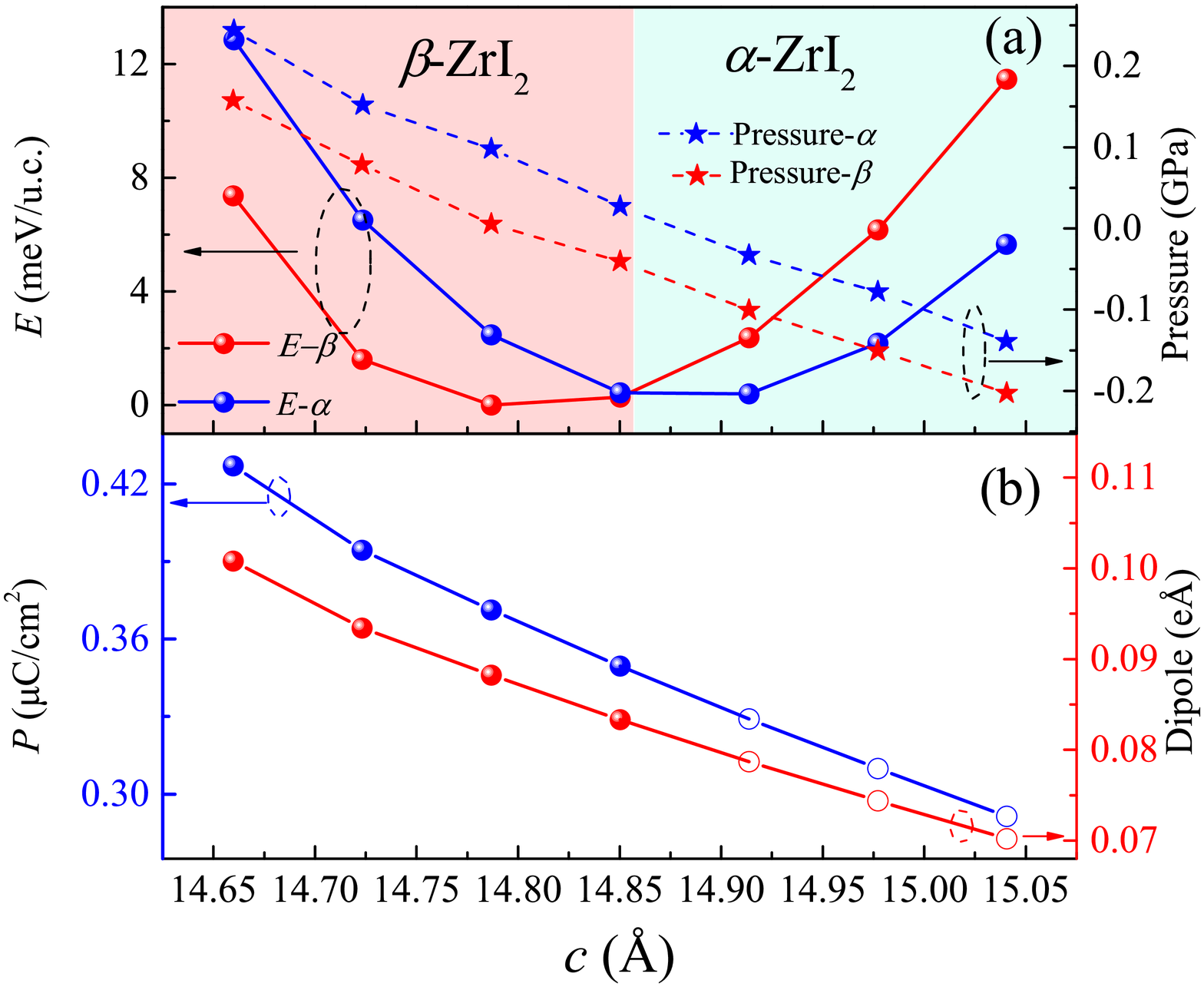}
\caption{The evolution of ZrI$_2$ under the uniaxial strain applied along the $c$-axis, as a function of the lattice constant of $c$-axis. (a) The compressive strain prefers the ferroelectric $\beta$-ZrI$_2$, while the pulling stress can lead to the $\alpha$-ZrI$_2$. The corresponding pressures of strain are also shown (right axis). (b) The negative piezoelectricity, namely the out-of-plane polarization is enhanced by pressure along the $c$-axis. Both the net polarization (left axis) and the electric dipole of a u.c. (right axis) are enhanced by pressure. The open circles denote the meta-stable $\beta$-ZrI$_2$.}
\label{fig5}
\end{figure}

Second, $\beta$-ZrI$_2$ may be also sensitive to the strain due to its loose vdW layer. Here a uniaxial strain is applied along the $c$-axis. The energy curves of $\beta$-ZrI$_2$ and $\alpha$-ZrI$_2$ as a function of the lattice constant $c$ have a crossover at $c=14.85$ \AA{} (i.e., $\sim+0.35\%$ from the optimized $\beta$-ZrI$_2$), as shown in Fig.~\ref{fig5}(a), which can be easily realized via $0.038$ GPa tensile stress. In the opposite direction, the compressive strain can further stabilize the $\beta$ phase and enhance its polarization, i.e., the negative longitudinal piezoelectricity [Fig.~\ref{fig5}(b)].

The enhancement of polarization upon the $c$-axis compression can be contributed by two factors: the first part is the reduced volume, especially for the shrunken soft vdW layer; the second part is the enhanced dipole of each u.c..
The first part was once proposed as the main contribution in CuInP$_2$S$_6$ and other layered ferroelectrics \cite{you2019origin,Qi-NPR-PRL-2021,PhysRevLett.119.207601}. However, more interestingly, here the electric dipole of $\beta$-ZrI$_2$ in each u.c. is also enhanced with decreasing lattice constant along the $c$-axis, as shown in Fig.~\ref{fig5}(b). This tendency is due to the unique origin of interlayer-sliding ferroelectricity, which is beyond the scenario of negative piezoelectricity in CuInP$_2$S$_6$. In other words, the closer neighboring ZrI$_2$ layers, the larger in-plane sliding, and the stronger out-of-plane polarization. In fact, for $\beta$-ZrI$_2$, the first part contributes only $\sim5\%$, while the second part contributes $\sim95\%$, to the negative longitudinal piezoelectricity.

The negative piezoelectricity is further confirmed by the DFPT calculation \cite{Lin-PRM-2019}. The elastic matrix of $\beta$-ZrI$_2$ has nine independent nonzero matrix elements ($C_{11}$, $C_{12}$, $C_{13}$, $C_{22}$, $C_{23}$, $C_{33}$, $C_{44}$, $C_{55}$, $C_{66}$) due to the space group $Pmn2_1$ \cite{PhysRevB.90.224104}. The piezoelectric tensor matrix has five independent matrix elements ($e_{15}$, $e_{24}$, $e_{31}$, $e_{32}$, $e_{33}$) for its point group $mm2$ \cite{deJong2015}. Then, the piezoelectric strain coefficient $d_{ij}$ can be calculated as:
\begin{equation}
d_{ij}=\sum_{k=1}^6 e_{ik}S_{kj},
\end{equation}
where $S=C^{-1}$ is the tensor for elastic compliance coefficients. The calculated piezoelectric $d_{33}$ is $-1.445$ pC/N, indeed a negative one. The values of more details of $C_{ij}$, $e_{ij}$, and $d_{ij}$ can be found in Table S2 of SM \cite{sm}.

\begin{figure}
\includegraphics[width=0.48\textwidth]{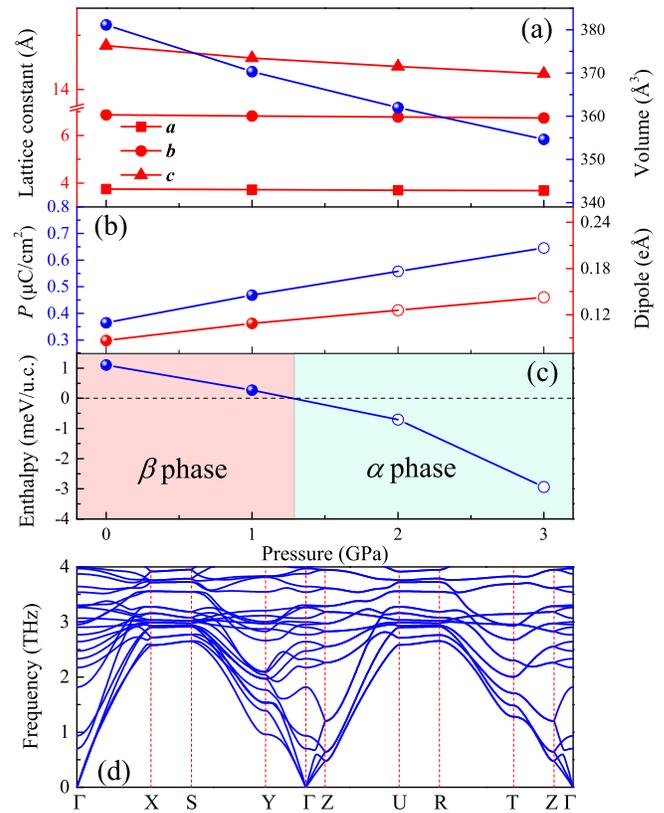}
\caption{The effects of hydrostatic pressure to $\beta$-ZrI$_2$. (a) The lattice constants (left axis) and the volume (right axis). (b) The polarization (left axis) and electric dipole of a u.c. (right axis). (c) The enthalpy difference between $\alpha$-ZrI$_2$ and $\beta$-ZrI$_2$ as a function of pressure, where the enthalpy of $\beta$-ZrI$_2$ is taken as the reference. $\alpha$-ZrI$_2$ becomes more stable above $\sim1.3$ GPa. (d) Phonon spectrum for $\beta$-ZrI$_2$ at $3$ GPa, which remains dynamically stable.}
\label{fig6}
\end{figure}

Third, the phase competition can be also tuned by hydrostatic pressure. Figure~\ref{fig6}(a) plots the lattice constants as functions of the hydrostatic pressure. As expected, the in-plane stiffness (along the $a$- and $b$-axes) is higher than the out-of-plane one (along the $c$-axis). For example, at $1$ GPa, the lattice constant $c$ decreases for $1.5\%$, while it only decreases $0.6\%$ for $a$, and then the total volume is reduced for $2.8\%$.

Similarly to aforementioned strain case, the polarization is also promoted by increasing hydrostatic pressure, as shown in Fig.~\ref{fig6}(b). Noting for most ferroelectrics, the hydrostatic pressure prefers to suppress their polarizations \cite{PhysRevLett.119.207601}. Not only the net polarization, but also the dipole of ZrI$_2$ in a u.c. is enhanced by hydrostatic pressure, as a characteristic result of interlayer-sliding ferroelectricity.

Since the volume of $\beta$ phase is larger than that of $\alpha$ phase \cite{corbett1982second}, the phase transition from $\beta$-ZrI$_2$ to $\alpha$-ZrI$_2$ may be triggered by compression. Figure~\ref{fig6}(c) plots the enthalpy difference between $\beta$-ZrI$_2$ and $\alpha$-ZrI$_2$ as a function of pressure. The $\alpha$ phase becomes more stable than the $\beta$ phase when the pressure is above $1.3$ GPa. Anyway, it is worth noting that $\beta$-ZrI$_2$ remains dynamically stable under moderate pressure. For example, there is no imaginary vibration mode in the phonon spectrum of $\beta$ phase at $3$ GPa, as shown in Fig.~\ref{fig6}(d). Therefore, this pressure-driving phase transition should be in the first-order and the ferroelectric $\beta$-ZrI$_2$ with enhanced polarization can co-exist with $\alpha$-ZrI$_2$ as the meta-stable phase under pressure.

\section{Conclusion}
In summary, based on the first-principles calculations, the structure and electric properties of vdW bulks ($\alpha$-ZrI$_2$ and $\beta$-ZrI$_2$) have been studied systematically. They are sister phases derived from the parent phase $s$-ZrI$_2$, and can coexist in real materials due to their proximate energies. Our calculation revealed a moderate out-of-plane polarization ($0.39$ $\mu$C/cm$^2$) for $\beta$-ZrI$_2$, resulting from the unbalanced Zr-I bonds due to interlayer sliding. The possible ferroelectric switching process is also nontrivial, mediated via its sister phase $\alpha$-ZrI$_2$, which leads to the barrier with double peaks.

In addition, the ferroelectric properties of $\beta$-ZrI$_2$ can be tuned effectively by mechanic approaches, including shear strain, uniaxial strain along the $c$-axis, as well as the hydrostatic pressure. The most exciting result is that the polarization can be increased by pressure and compressive strain, i.e., the negative piezoelectricity. Such anomalous effect is directly associated with its vdW structure and its interlayer sliding ferroelectricity.

\textit{Note.} Very recently, Ma \textit{et al.} independently studied the ferroelectric properties of $\beta$-ZrI$_2$ \cite{ma-ZrI2-arXiv-2021}. Despite the common base of ferroelectricity, our work and Ma's work contain different contents: we focused on the mechanical tuning and negative piezoelectricity, while they focused on the ferroelectric domains.

\begin{acknowledgments}
We thank Yang Zhang and Shan-Shan Wang for useful discussions. This work was supported by the National Natural Science Foundation of China (Grant No. 11834002). We thank the Tianhe-II of the National Supercomputer Center in Guangzhou (NSCC-GZ) and the Big Data Center of Southeast University for providing the facility support on the numerical calculations.
\end{acknowledgments}

\bibliography{apssamp}
\bibliographystyle{apsrev4-2}
\end{document}